\documentclass[3p,times, twocolumn]{elsarticle}

\usepackage{ecrc}

\volume{00}

\firstpage{1}

\journalname{Acta Materialia}

\runauth{}

\jid{procs}

\jnltitlelogo{Acta Materialia}

\CopyrightLine{2011}{Published by Elsevier Ltd.}

\usepackage{amssymb}
\usepackage[figuresright]{rotating}
\usepackage{siunitx}

\begin{document}

\begin{frontmatter}


%%\title{Title\tnoteref{label1}}

\dochead{}
%% e.g. \dochead{17th International Conference on Dynamical Processes in Excited States of Solids}

\title{In situ visualization of Ni-Nb bulk metallic glasses phase transition.}


\author[1]{A.I. Oreshkin} 
\author[1]{V.N. Mantsevich} 
\author[1]{S.V. Savinov}
\author[1,2]{S.I. Oreshkin}
\author[1]{V.I. Panov}
\author[3,4,5]{A.R. Yavari}
\author[6]{D.B. Miracle}
\author[3]{D.V. Louzguine-Luzgin}
\address[1]{Department of  Physics, Moscow State
University, 119991, Moscow, Russia}
\address[2]{Moscow, Sternberg Astronomical Institute, Moscow State
University, 119991, Russia}
\address[3]{WPI-AIMR, Tohoku University 2-1-1 Katahira, Aoba-ku,
Sendai 980-8577 Japan}
\address[4]{SIMAP-LTPCM, Institut National Polytechnique de Grenoble, St-Martin-d'Hères Campus, Grenoble, BP 75, 38402, France}
\address[5]{European Synchrotron Radiation Facility, 38042, Grenoble, France}
\address[6]{Materials and Manufacturing Directorate, Dayton, OH USA}

\begin{abstract}
We report  the results of \textit{in situ} investigation of the Ni-based bulk metallic glass structural evolution and crystallization behavior.
The X-ray diffraction (XRD), transmission electron microscopy (TEM), nano-beam diffraction (NBD), differential scanning
calorimetry (DSC), radial distribution function (RDF) and scanning probe
microscopy/spectroscopy (STM/STS) techniques were applied to analyze the
structure and electronic properties of $Ni_{63.5}Nb_{36.5}$ glasses before and
after crystallization. It was proved that partial surface crystallization of
$Ni_{63.5}Nb_{36.5}$ can occur at the temperature lower than that for the full sample
crystallization. According to our STM measurements the primary crystallization
originally starts with the $Ni_{3}Nb$ phase formation as a leading eutectic phase. It was shown that
surface crystallization drastically differs from the bulk crystallization  due
to the possible surface reconstruction. The mechanism of $Ni_{63.5}Nb_{36.5}$
glass alloy 2D-crystallization was suggested, which corresponds to the
local metastable $(3\times3)-Ni(111)$ surface phase formation. The possibility
of different surface nano-structures development  by the annealing of the
originally glassy alloy in ultra high vacuum at the temperature lower, than the
crystallization temperature was shown. The increase of the mean square surface
roughness parameter $R_{q}$ while moving from glassy to fully crystallized state
can be caused by concurrent growth of $Ni_{3}Nb$ and $Ni_{6}Nb_{7}$ bulk phases.
The simple empirical model for the estimation of $Ni_{63.5}Nb_{36.5}$ cluster
size was suggested, and the obtained value (about 8 \AA) is in good
agreement with the STM measurements data (8 \AA-10 \AA).
\end{abstract}


\end{frontmatter}

% \linenumbers

\section{Introduction}
\label{1}
Amorphous metallic glasses were segregated in a separate class of materials
starting from classic rapid-solidification experiments on Au–Si alloys performed by
Duwez and colleagues \cite{Klement}. Later a new class of metallic glasses
called bulk metallic glasses with high glass-forming ability were invented
~\cite{Inoue, Johnson}. Metallic glasses (especially bulk metallic glasses)
currently attract a significant attention of the scientists involved in the
materials science research ~\cite{Inoue, Johnson, Greer, Tian, Zeng, Ye}.
In these works the unique mechanical (high strength and large elastic
deformation limit), chemical (high corrosion resistance) and magnetic properties
(either extremely soft or moderately hard magnetism) were intensively studied.
The physical and chemical properties of metallic alloys in  an amorphous state
significantly differ from the properties of metallic alloys in a crystalline
state. The strength and hardness values of the amorphous alloys exceed that of the
crystals one, due to ``disordered`` local structure of metallic glasses, while the
elastic modulus of normal elasticity is slightly lower than in the crystalline state. An elastic
modulus difference can be caused by the more loosely packed structure of metallic
glasses due to the smaller number of the nearest neighbors in the atomic
structure. Therefore the creation of partially crystallized structures can open
new outlooks for metallic glasses properties improvement ~\cite{Louzguine}.

Until now the majority
of the investigations dedicated to metallic glasses study  were oriented on the
obtaining of the information about local arrangement of the constituted
components. Transmission electron microscope and diffraction methods were mainly
applied for these purposes ~\cite {Louzguine-Luzgin}. Spectroscopic information
was obtained from photoemission experiments ~\cite{Rotenberg}. But a detailed
information on a subnanometer scale is required for full understanding of
processes which take place during metallic glasses phase transition from
the amorphous state to crystalline one. Scanning tunneling microscopy/spectroscopy
(STM/STS) \cite{Binnig, Rohrer} is a very powerful method for getting the
necessary information, but only a few works ~\cite{Oreshkin, Oreshkin1,
Ashtekar} were devoted to STM/STS study of bulk metallic glasses surface. In
~\cite{Ashtekar} scanning tunneling microscopy was applied to investigate time
evolution  of the metallic glassy surface  with temporal resolution as fast as 1 minute per scan
and extending up to 1000 minutes. It was shown that rearrangements of surface
clusters occur almost exclusively by two-state hopping. In ~\cite{Oreshkin} UHV
STM/STS measurements of bulk metallic glasses surface topography and electronic
properties at the room temperature were performed. The pseudo-gap at Fermi
energy and linear part of the normalized tunneling conductivity spectra were
revealed in consistence with suggested theoretical model. It was shown that the
additional features on the averaged normalized tunneling conductivity are
connected with the presence of localized states in individual clusters on the
$Ni_{63.5}Nb_{36.5}$ surface. The theoretical analysis of electron elastic
scattering on the impurities inside each cluster and on the random defects on
the cluster boundaries in the presence of intra-cluster Coulomb interaction of the
scattering electrons has been applied to explain the experimental results. 
In ~\cite{Oreshkin1} we investigated the processes of bulk metallic glasses
crystallization by means of STM. We pointed out the existence of three different
surface lattice structures observed during full $Ni_{63.5}Nb_{36.5}$ crystallization
that are completely different from those formed according to Ni-Nb binary phase
diagram in the bulk  sample.

In this work we present the results of the comprehensive investigation of the
Ni-based bulk metallic glass structural evolution and crystallization 
behavior by X-ray diffraction (XRD), including radial
distribution function (RDF) analysis, transmission electron microscopy (TEM),
nano-beam diffraction (NBD), differential scanning calorimetry (DSC)  and scanning probe microscopy/spectroscopy (STM/STS)
methods.

\section{Experimental procedure}

An ingot of the $Ni_{63.5}Nb_{36.5}$ alloy (composition is given in nominal
atomic percentage) was prepared by arc-melting mixtures of Ni (99.99 mass.\%
purity) and Nb (99.9 mass.\% purity) under an argon atmosphere. From this ingot,
bulk rod sample of 1 mm in diameter was prepared by the injection casting
technique in an argon atmosphere from the initial arc-melted ingot. The
structure of the cast samples was examined by conventional X-ray diffractometry
(XRD).

The samples were cut along the rod  axis before loading to the STM
holder. The surface suitable for STM/STS study was obtained by grinding process
with further treatment of the surface using polishing slurry with particle sizes
of 5, 1, and 0.5 $\mu m$ to get flat and mirror like surface. The quality of
prepared surface was controlled by optical microscope with 800 nm resolution.
Before placing the samples in STM holder they were ultrasonically cleaned in
acetone and distilled water. Samples were degassed at 673 K during 24 hours and
processed with argon-ion-sputtering (1.5 keV, 30 $\mu$A, 60 min) at
$9.0\times10^{-6}$ torr argon pressure. At the final stage of clean surface
preparation the samples were heated again starting from 673 K (to get clean
surface in amorphous state) (which is 257 K  below its crystallization
temperature equal to 930 K ~\cite{Xia}) and higher (up to 973 K) during 12 hours
at $10^{-10}$ torr pressure range.

All STM/STS experiments were performed using
commercial UHV Omicron system with base pressure $4\times10^{-11}$ torr at
room temperature. In our experiments we used tungsten tips obtained by
electrochemical etching. To remove the oxide layer from the STM tip apex, they
were annealed at 1473 K at ultra high vacuum conditions (less then
$1.0\times10^{-10}$ torr). Bias voltage was applied to the sample, while STM tip
was virtually ground. The current imaging tunneling spectroscopy (CITS)  curves
were obtained simultaneously with constant current images. The STM images were
400$\times$400 points in size, while the CITS images were 80$\times$80 points.
The I(V) spectrum was taken at each fifth point of each fifth line of the
scanning frame.

Transmission electron microscopy (TEM) investigations were
carried out using a JEOL JEM 2010 microscope operating at 200 keV. The samples for
TEM were prepared mechanically (down to 10 $\mu m$ thickness) and  by subsequent
ion polishing  (down to electron-beam transparency).  In order to avoid
structural damage the ion-beam energy was kept as low as 2.0 keV.  Nanobeam
diffraction (NBD) patterns were also obtained in the transmission
electron microscope.

Synchrotron radiation x-ray diffraction measurements 
were carried out using a high energy monochromatic beam at ID11 beam line of the
European Synchrotron Radiation Facility (ESRF) equipped with a nitrogen cooled
double-silicon monochromator. The photon energy was 94 keV. After correction
for air scattering, polarization, absorption ~\cite{Wagner}, and Compton
scattering ~\cite{Cromer}, the measured intensity was converted to electron
units per atom with the generalized Krogh–Moe–Norman method ~\cite{Wagner1},
using the x-ray atomic scattering factors and anomalous dispersion 
corrections ~\cite{Ibers}. The total structure factor S(Q) (Q-scattering vector)
and the interference function Qi(Q) were obtained from the coherent scattering
intensity by using atomic scattering factors ~\cite{Waseda}. The values of Qi(Q)
less than 18 $nm^{-1}$ were smoothly extrapolated to Q = 0. The radial
distribution (RDF(r)) and pair distribution (PDF(r)) functions were obtained by
the Fourier transformation of Qi(Q).

\section{Results and discussion}

\begin{figure}[hbtp]
\centering
\includegraphics[width=80mm]{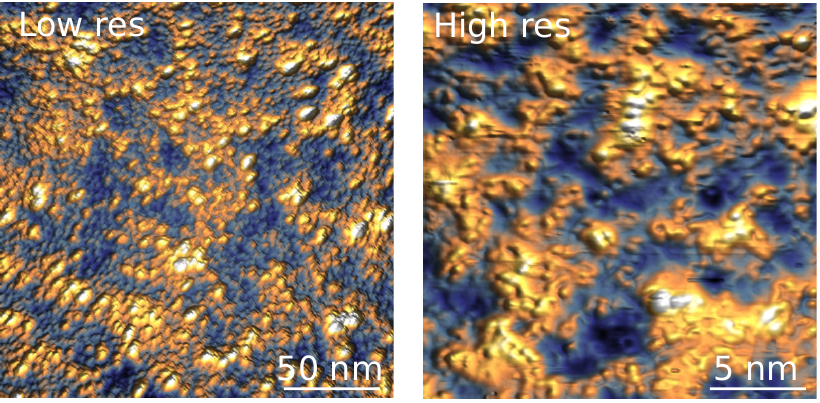}
\caption{\label{Fig1} (a) STM images of the  $Ni_{63.5}Nb_{36.5}$ sample in
amorphous  state; $200nm\times 200 nm$; V=-2.0 V, I=40 pA, (b) STM image of the
$Ni_{63.5}Nb_{36.5}$ sample in amorphous state, $20nm\times 20 nm$; V=-2.1 V,
I=103 pA. Note the difference in average cluster size.}
\end{figure}

Fig.1 demonstrates typical STM images of the $Ni_{63.5}Nb_{36.5}$  metallic
glass surface in the amorphous state. The procedure of surface preparation was
described in the previous section. After ion bombardment the sample was heated
up to 673 K. The  high resolution STM (Fig.~\ref{Fig1}b) image shows surface
structure consisting of  8--10 \AA~ size clusters.  Small clusters on a
surface  have a tendency to coalesce forming the clusters with typical size of
about 50 \AA  ~\citep{Oreshkin} (Fig.~\ref{Fig1}a).   We were not able to
observe any ordered structure in  STM or HRTEM images. The obtained results demonstrate
the absence of long-range order on the surface. We suppose that the sample
surface can be treated as a disordered structure formed by set of random
clusters. 

\begin{figure}[hbtp]
\centering
\includegraphics[width=75mm]{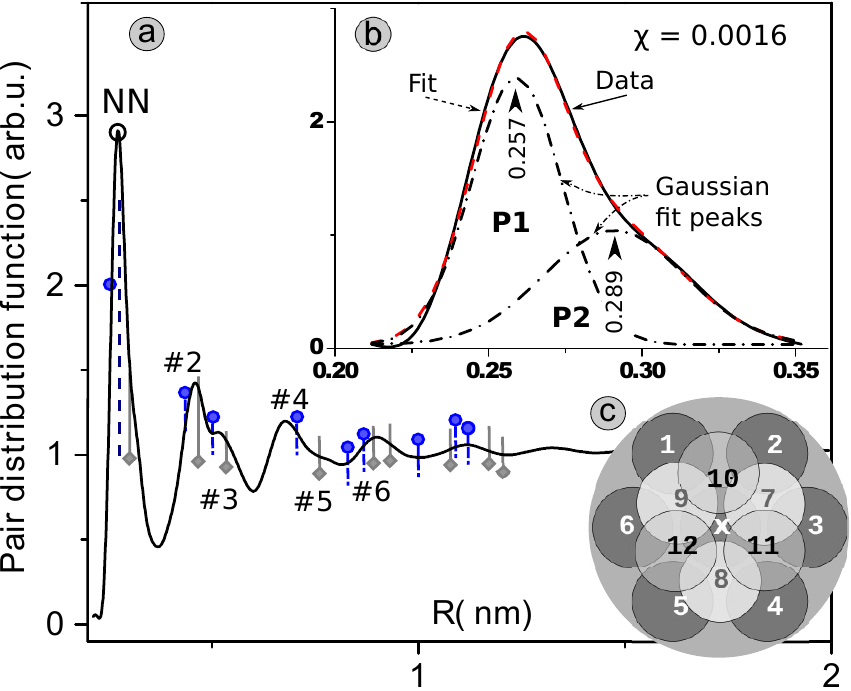}
\caption{\label{Fig2} a) Normalized radial distribution function for
$Ni_{63.5}Nb_{36.5}$  sample in amorphous state. (b) demonstrates the
thin structure of the first visible peak.
\newline
The inset (b) shows deconvolution of the first peak into Ni-Ni and Nb-Nb components.
Peak locations and intensities predicted from the efficient cluster packing
model are shown for Ni-Ni and Nb-Nb  separations. The predicted location
and intensity of the Nb-Ni 1st neighbor peak is also shown.
\newline
(c) Schematic model of individual cluster observed on $Ni_{63.5}Nb_{36.5}$
surface in the amorphous state. Atoms 10, 11, 12 are located in the upper layer,
x, 1, 2, 3, 4, 5, 6 in the middle layer and 7, 8, 9 in the bottom layer. The
distance between the atom "x" and any neighbor atoms (1--12) is 2.57 \AA. The
outer surround corresponds to the average size of cluster observed in STM
experiments.}
\end{figure}

To get information about short-range and medium-range ordering we derived the
radial distribution function (Fig.~\ref{Fig2}) from a synchrotron XRD spectrum. The efficient cluster packing
(ECP) model ~\cite{Miracle1, Miracle2, Miracle3} was used to estimate the
location and intensity of the RDF
peaks, giving a more direct structural interpretation. Both fcc and hcp cluster
packing give essentially equivalent peaks and intensities, fcc cluster packing
is used for the predictions here. Contact between like atoms is unavoidable in
this solute-rich glass, and so the nearest neighbor (NN, or 1st) peak includes
both Ni-Ni and Nb-Nb contributions, shown as P1 and P2 respectively in the inset
in Figure~\ref{Fig2}. The partial coordination numbers are estimated from
the ECP model ~\cite{Miracle4} to be ${\mathrm {Z_{NiNi} = 7.56, Z_{NiNb} = 5.42,
Z_{NbNi} = 11.1, Z_{NbNb} = 3.61}}$, so that the dominant contribution to the NN
peak comes from Nb-Ni contacts, also shown in Figure 2.

The location and intensities of the 2nd and 3rd peaks at about 4.6 \AA~ and 5.2~
\AA~ are well-predicted, and represent the spacings between clusters that
overlap in the 1st coordination shells. The 2nd and 3rd peaks represent atoms on
the $\gamma$ and $\beta$ sites of the ECP model, respectively. Peaks \#4 and \#5
at roughly 6.8 \AA~ and 7.7~ \AA~ represent the closest contact between
non-overlapping clusters along $<110>$ directions of the cluster unit cell.
The difference between these two peaks occurs from the difference between Ni-Ni
separations (\#4) and Nb-Nb separations (\#5). Peak \#6 is the next-nearest
separation between non-overlapping clusters and occurs along $<111>$
directions of the cluster unit cell. The ECP model does not allow estimation of
Nb-Ni separations beyond nearest neighbor contact, but the agreement between
prediction and experiment is still reasonably good out to distances of about 1
nm. 

The position of the first peak ($P_{1}$)  (inset in the Fig.~\ref{Fig2}) allows
us  to estimate the nearest neighbor  distance (2.57 ~\AA~in our case). To our
opinion, the distances between Ni and Nb atoms ($R_{Ni-Nb}=2.72$ ~\AA)  and
between  two nearest Ni atoms ($R_{Ni-Ni}=2.5$ ~\AA) can contribute to the first
peak. The second peak ($P_{2}$) is located at 2.89 ~\AA~ and can be responsible
for the distance between two nearest Nb atoms ($R_{Nb-Nb}=2.94$ ~\AA). The area
of the first maximum gives evidence of the  nearest neighbors number  or
coordination number ''Z". We can conclude that in our case the coordination
number is equal to 12.47, taking the integral over the RDF(R) curve in the
range of values giving contribution for the first peak  from 0.22 to 0.35 nm (this means that the
surface contains the equal number of clusters with 12 nearest neighbors (13
atoms)). Fig.~\ref{Fig2} demonstrates a
model of one cluster consisting of 13 atoms (atom, marked by "x", has 12
neighbors). The distance between atom "x" and any of neighbor
atoms  is 2.57 ~\AA. In our model we used the values of 1.25 ~\AA~and 1.47 ~\AA~for
the radii of  Ni and Nb atoms correspondingly (Goldschmidt metallic radii) ~\cite{Smithells}. In such a way
one can estimate the minimum  cluster size (all of the cluster atoms are Ni
atoms) and the maximum size (all of the cluster atoms are Nb atoms). The
obtained values (7.64 ~\AA~and 8.08 ~\AA) are in the good agreement with STM
measurements (8--10 ~\AA).

Here we have to note that this is just an estimation.  More accurate
approximation can be reached with DFT modeling. Anyway, spatial size estimation
error  should be  relatively small due to the fact that  local electronic
density of states distribution for metal can be treated as simple superposition
of metallic radii spheres. 

\begin{figure}[hbtp]
\centering
\includegraphics[width=80mm]{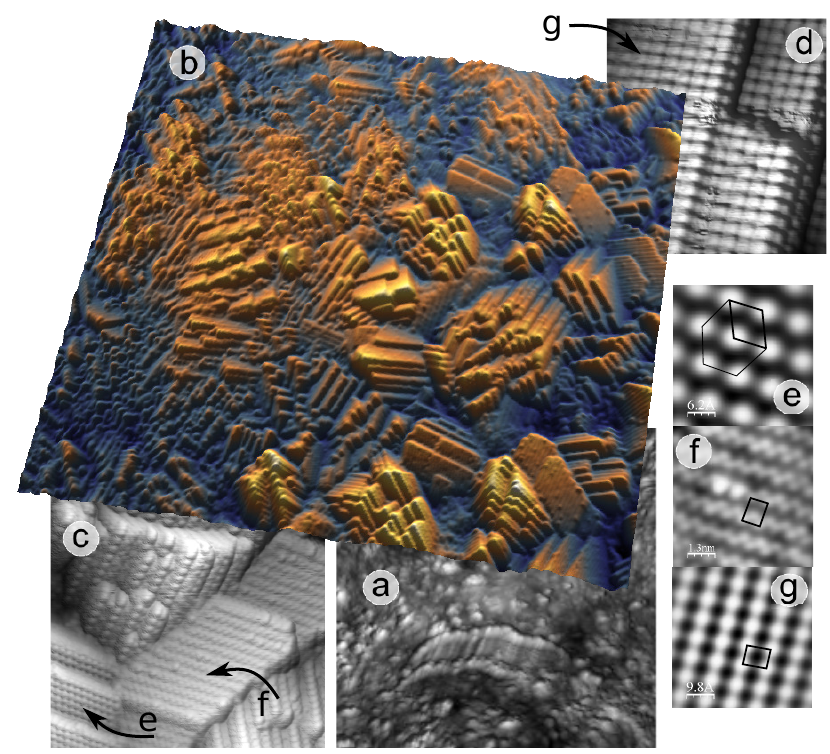}
\caption{\label{Fig3} (a) Constant current STM image of $Ni_{63.5}Nb_{36.5}$
surface at the initial stage of crystallization. The $Ni_{63.5}Nb_{36.5}$ sample
was annealed at 883 K; $100 nm\times 100 nm$; V=-2.0 V; I=67 pA; (b)  STM image
of the $Ni_{63.5}Nb_{36.5}$ sample annealed at 973 K representing a
nanostructured eutectic; $150 nm\times 150 nm$; V=-2.1 V; I=62 pA; (c)  STM
image of the $Ni_{63.5}Nb_{36.5}$ sample annealed at 973 K representing a
nanostructured eutectic; $30 nm\times 30 nm$; V=-2.1 V; I=62 pA;  (d)  STM image
of the $Ni_{63.5}Nb_{36.5}$ sample annealed at 973 K representing a part of
surface area observable in STM image Fig.~\ref{Fig3} (a); $15 nm\times 15 nm$; V=-1.8 V;
I=105 pA ; (e) High resolution STM image of the area indicated by arrows e in
Fig.~\ref{Fig3} c: $3.1\times 3.1 nm$; V=-2.1 V; I=62 pA; (f) High resolution STM image of
area indicated by arrows f in Fig.~\ref{Fig3} c: $6.5\times 6.5 nm$; V=-2.1 V; I=62 pA; (g)
High resolution STM image of the area shown in Fig.~\ref{Fig3} d and indicated by arrows g:
$4.9 nm\times 4.9 nm$; V=-2.1 V; I=105 pA. }
\end{figure}

To analyze the initial phase of $Ni_{63.5}Nb_{36.5}$ metallic glass 
crystallization the samples were annealed step by step starting from  883 K,
which is still lower than the crystallization temperature (955 K at 40 K/min.) up to 973 K.
The surface crystallization starts to appear even  at lowest temperature of 883 K
(Fig.~\ref{Fig3}a), i.e.  at temperature noticeably lower than the bulk eutectic
one ~\cite{Luborsky}. This phenomenon can be connected with the surface-induced
crystallization ~\cite{Koster}.

The melting point of Ni (1726 K) is almost two times smaller than the melting
point of Nb (2741 K). If a metallic glass  has the local volumes enriched by
Ni, the Ni atoms can gain  additional mobility in these areas  at the
temperatures significantly lower than crystallization temperature. At 900 K the
Nb self-diffusion coefficient is seven times lower than that for Ni
~~\citep{Smithells}. In this case the formation of  nano-crystallites
from one of  amorphous glass components is  possible (Ni nano-crystals, for
example). The increase of both the annealing time and the temperature leads to
the local formation of eutectic phases ($Ni_{3}Nb$ and $Ni_{6}Nb_{7}$ in our case).
We suppose that $Ni_{3}Nb$ and $Ni_{6}Nb_{7}$ phases are growing concurrently. 
Atomic planes of a few atomic layers thick coming out at the surface (Fig.~\ref{Fig3}b) 
most probably represent  eutectic structure, composed of the phases including a different Ni content.
 
\begin{figure}[hbtp]
\centering
\includegraphics[width=60mm]{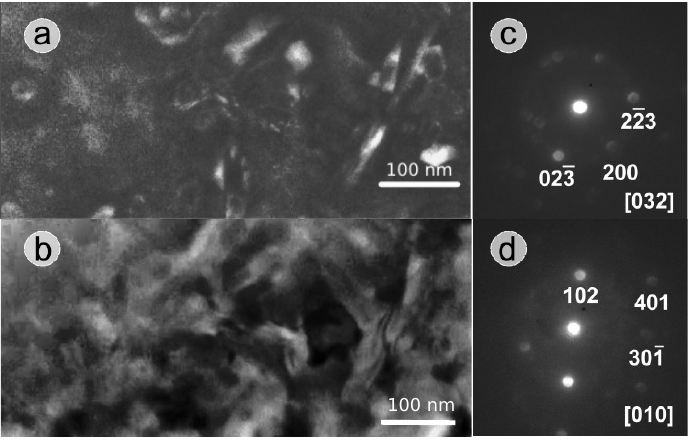}
\caption{\label{Fig4}Dark-field (a) and bright-field (b) TEM images of the
sample annealed at 973 K during 120 seconds represent a nanostructured eutectic.
(c) nanobeam diffraction patterns obtained from $Ni_{3} Nb$ phase in the
eutectic mixture. (d) nanobeam diffraction patterns obtained from $Ni_{6}Nb_{7}$
phase in the eutectic mixture. Zone axes are given in square brackets.}
\end{figure}

The crystallization of the $Ni_{63.5}Nb_{36.5}$ bulk metallic glass was also studied by TEM.
Dark-field and bright-field images of the $Ni_{63.5}Nb_{36.5}$ sample annealed
at 973 K are shown in Fig.~\ref{Fig4}(a, b) correspondingly. Nano-structured
eutectic is visible in Fig.~\ref{Fig4}.  The results obtained from Nano-Beam
Diffraction (NBD) measurements clearly demonstrate an existence of both
mentioned above $Ni_{3}Nb$ and $Ni_{6}Nb_{7}$ phases in the bulk of our sample
(Fig.~\ref{Fig4} (c,d)).

\begin{figure}[hbtp]
\centering
\includegraphics[width=80mm]{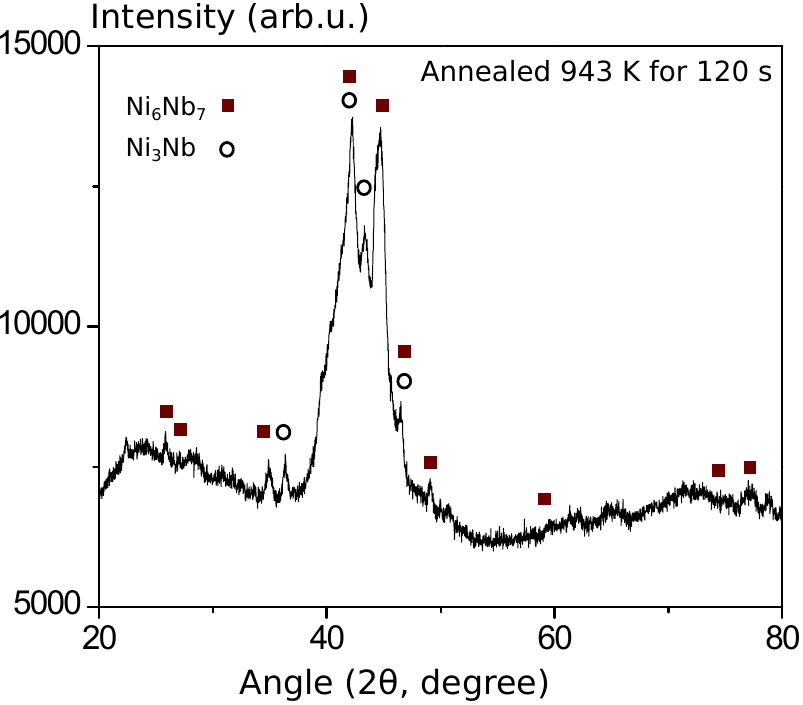}
\caption{\label{Fig5} XRD pattern recorded for the initial crystallization of
$Ni_{63.5}Nb_{36.5}$ sample.}
\end{figure}

Figure~\ref{Fig5} demonstrates XRD pattern of the initial crystallization of
$Ni_{63.5}Nb_{36.5}$ metallic glass sample. There are several well
distinguishable peaks visible on the diagram  corresponding to  $Ni_{3}Nb$
and $Ni_{6}Nb_{7}$ phases marked as rectangles and circles
correspondingly.

\begin{figure}[hbtp]
\centering
\includegraphics[width=75mm]{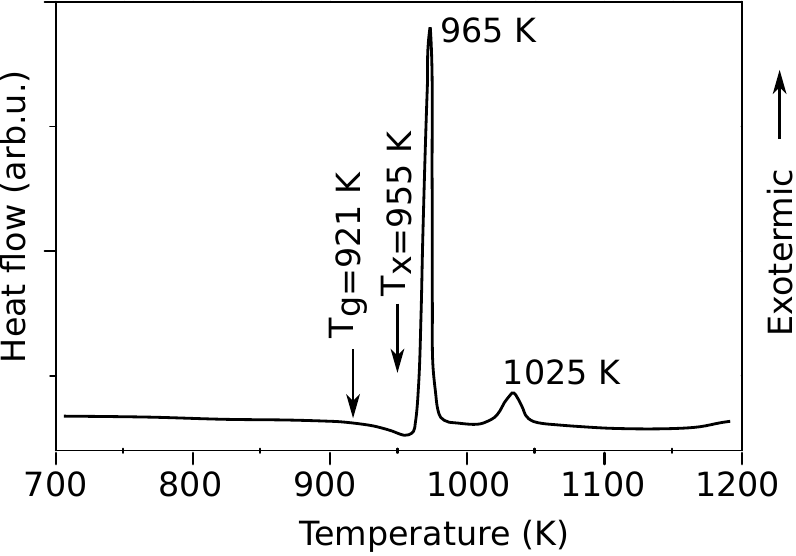}
\caption{\label{Fig6} DSC curve of $Ni_{63.5}Nb_{36.5}$ alloy at a heating rate
of 0.67 K/s under a flow of purified argon.}
\end{figure}

Figure~\ref{Fig6} shows the DSC curve of $Ni_{63.5}Nb_{36.5}$
alloy. The alloy exhibits an endothermic like event, characterized by a glass
transition temperature $T_{g}=921 K$. This temperature is a starting point for
the amorphous phase-liquid transition. From the temperature $T_{x}=955 K$ the
eutectic transformation starts. The second maximum is likely responsible for the growth of
eutectic phases.

Here we must note that all of the applied methods: TEM, NBD, XRD and DSC are
very useful  for the analysis of amorphous alloys initial crystallization but the
information extracted from these experiments mainly regards to the bulk
properties. Even TEM can be characterized as a bulk method despite of small
samples thickness. Besides it is a diffraction method, producing a picture in the 
reciprocal space. The processes taking place at amorphous alloys surface during
initial crystallization can  differ radically from the bulk ones. 

Another interesting point is the origin of the crystallization phenomenon. Does
the crystallization process starts to develop from the  surface to bulk or vice
versa? To get the answers for this question the STM/STS measurements were performed.
The undeniable advantage of STM method is a very high spatial resolution. The
results of $Ni_{63.5}Nb_{36.5}$ STM investigations are presented in
Figure ~\ref{Fig3}. Fig.~\ref{Fig3}b shows the  STM image of the
$Ni_{63.5}Nb_{36.5}$ sample  annealed at 973 K. The STM image has the size which
approximately corresponds to the size of nano-crystal, embedded in the amorphous
matrix (bright spot in Fig.~\ref{Fig4}b), and consists of many randomly oriented facets. 
The angular distribution
for plane inclinations reveals that the angle $\theta$ for the most often
observed facets appears in the range 2--12$^\circ$. Here $\theta$ is the angle
between the normal vectors to facet and to horizontal plane, this means that $\theta$ =
0 for horizontal facets and it
increases with the slope. 

Fig.~\ref{Fig3} c, d  demonstrates a part of area presented in Fig.~\ref{Fig3}b
with higher resolution. The analysis of the facets structure reveals the presence of
three different surface reconstructions: hexagonal, rectangular and zigzag
(Fig.~\ref{Fig3} (e, f, g). It is possible to observe three surface structures
on different image areas marked by the corresponding arrows. STM image in
Fig.~\ref{Fig3}e shows well defined hexagonal structure with rhomb-shaped  unit
cell with sizes 7.6\AA~$\times$~7.6\AA~ (marked by  black lines in
Fig.~\ref{Fig3}e). Figure ~\ref{Fig3}f demonstrates zigzag surface
structure which takes place in the area indicated by arrows \textbf{f} in
Fig.~\ref{Fig3}c. The rectangle shows the primitive unit cell with sizes
7.9\AA~$\times$~10.3\AA. Fig.~\ref{Fig3}g is the high-resolution STM image of
rectangular surface structure observed in the area indicated by arrow \textbf{g}
in Fig.~\ref{Fig3}d. The determined parameters of the  unit cell are 8.4\AA~
$\times$~7.6\AA. This one is identical to the structure observed in our
experiments with the partially crystallized $Ni_{63.5}Nb_{36.5}$ metallic glass
(Fig.~\ref{Fig3}a). Most probably it corresponds to the orthorhombic $Ni_{3}Nb$
crystal system class  due to the same symmetry of unit cell.   

All the STM images were obtained at negative bias voltage (in filled states bias
range). No one of the surface structures corresponds to possible periodicity of
$Ni$ monocrystal, $Nb$ monocrystal or $Ni_{3}Nb$, $Ni_{6}Nb_{7}$ crystals.
Because of tunneling spectra similarity we suggest that the rectangular and
zigzag structures observed in our STM/STS experiments are two possible surface
reconstructions of the $Ni_3Nb$ phase~\citep{Oreshkin1}. The best correspondence
was found for hexagonal structure and metastable surface reconstruction
$(3~\times~3)-Ni(111)$. It is well known that the free-electron-like surface
states, the Shockley surface states (SS), are present on the Ni(111) surface
~\cite{Magaud, Nishimura, Braun}.  If the observed in our experiments hexagonal
surface structure (Fig.~\ref{Fig3}e) corresponds to surface reconstruction
$(3~\times~3)-Ni(111)$, then  the  Shockley surface states  should  be visible
in our scanning tunneling spectra. 

\begin{figure}[hbtp]
\centering
\includegraphics[width=80mm]{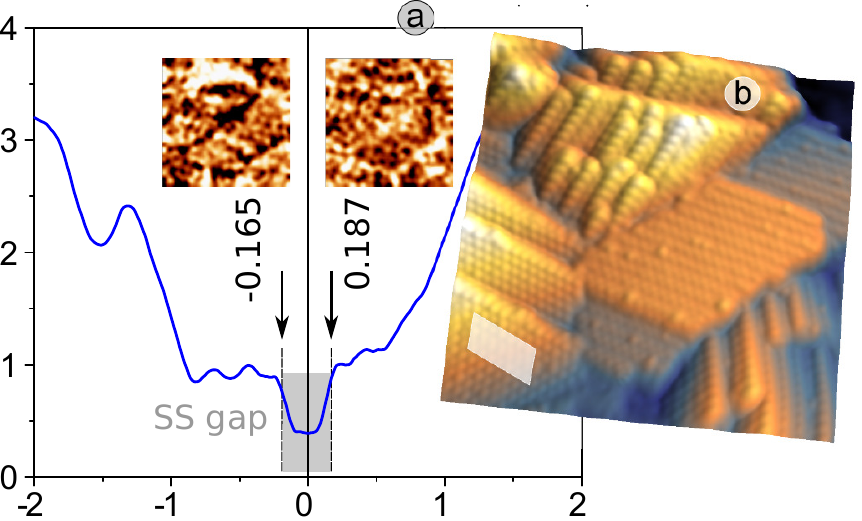}
\caption{\label{Fig7} (a) Average normalized tunneling conductivity spectrum.
The averaging was performed within the limits of rectangular shown in Fig.8b.
Additional frames inside of Fig.8a are the CITS images obtained at values of
bias voltage corresponding to maxima of conductivity curve indicated by arrows; 
(b) Constant current STM image; $30 nm\times 30 nm$; V=-2.1~V; I=62 pA}
\end{figure}

To verify this statement the current imaging  tunneling spectroscopy (CITS)
measurements were performed.  Fig.~\ref{Fig7}b is the constant current tunneling
image. CITS images inside the Fig.~\ref{Fig7}a were measured at -0.17 V and 0.19
V correspondingly.  They represent the same area of the surface as constant
current STM image in Fig.~\ref{Fig7}b with lower resolution ($80\times 80$
points). Averaged normalized tunneling conductivity spectrum is presented in
Fig.~\ref{Fig7}a. The averaging was performed within the area, containing
rectangular surface structure shown in Fig.~\ref{Fig7}b.

It was demonstrated in ~\citep{Oreshkin1} that the position of the peaks in the
normalized tunneling conductivity curve does not depend on the local point in
the range of chosen structure. Regarding to our STS  measurements above
hexagonal surface structure we consider that the normalized tunneling
conductivity curve  maxima at -0.165 V and 0.187 V bias voltage are due to
Shockley surface states below and above $E_{F}$ with an exchange splitting of
$\sim$ 352 meV. This value is in the good agreement with theoretical works that
have also predicted two surface states bands, which are separated by 180-350 meV gap
~\cite{Memmel, Braun1, Ohwaki, Lobo-Checa}.

Consequently the $Ni_{63.5}Nb_{36.5}$ crystallization process can be divided
into three stages (formation of quasi-ordered structures with 1-D periodicity,
2-D periodicity that are visible in Fig.~\ref{Fig3}b and the formation of 3-D
nanocrystals (Fig.~\ref{Fig5}, Fig.~\ref{Fig6})~\cite{Liu}. On the second stage
(2D-crystallization) one can observe  the formation of $Ni_{3}Nb$ phase
(rectangular and zigzag surface structures) and $(3~\times~3)-Ni(111)$ surface
reconstruction (metastable phase of Ni areas). The peak
on $Ni_{63.5}Nb_{36.5}$ alloy DSC curve (Fig.~\ref{Fig6}) located at 965 K most
probably corresponds to this stage. The second maximum at 1025 K likely represents
the grain growth process from nano- to micro-scale  size of $Ni_{3}Nb$ and $Ni_{6}Nb_{7}$ phases.

Using the parameter
$$R_{q}=\sqrt{\frac{1}{L}\int^{L}_{0} Z^{2}dx},$$

\noindent where L is the sample length [{\AA}], Z is the profile height along
selected direction [{\AA}], X is the distance, we estimated mean square roughness of the studied
sample in three different states (amorphous, partially crystallized and full
crystallized). Averaged values for amorphous, partially crystallized and full
crystallized states are $R_{q}=0.19; R_{q}=0.17; R_{q}=0.30$ correspondingly.
The $R_{q}$ reduction for partially crystallized surface in comparison with
amorphous state is
probably caused by material redistribution during crystallization. The surface
structure formed due to primary crystallization is atomically flat and uniform
and therefore induces the mean square roughness parameter reduction. The $R_{q}$
enlargement in full crystallized state is caused by surface deformation  during
concurrent bulk phases $Ni_{3}Nb$, $(3\times 3)-Ni(111)$ and $Ni_{6}Nb_{7}$
growth. 

\section{Conclusion} 
The direct nanoscale range visualization of the structural evolution on the bulk
metallic  glassy sample was performed by means of STM/STS,  X-ray diffraction,
transmission electron microscopy, nano-beam diffraction, differential scanning
calorimetry and radial distribution function methods. The typical size of the
grains in the bulk glassy alloy was estimated by direct STM visualisation. The
model for estimation the  minimal  and maximal cluster sizes of
$Ni_{63.5}Nb_{36.5}$  alloy  was suggested and given a good correspondence with the RDF(R).

It was shown that partial surface crystallization of $Ni_{63.5}Nb_{36.5}$ can
start at the temperature significantly lower than for the sample's bulk
crystallization. The mean square roughness $R_{q}$ enlargement for fully
crystallized state is caused by concurrent bulk phases $Ni_{3}Nb$ and
$Ni_{6}Nb_{7}$ growth.

It was shown that surface crystallization can drastically differ from the bulk
one due to possible surface reconstruction. Three different surface structures
(rectangular, zigzag and hexagonal ) were observed during  the sample
crystallization. It was proved that the formation of two equilibrium
$Ni_{3}Nb$ and $Ni_{6}Nb_{7}$ phases on the surface is a very complicated process which includes
the two additional steps, namely the formation of quasi-ordered structures with
1-D periodicity and 2-D periodicity. The mechanism of $Ni_{63.5}Nb_{36.5}$ alloy
2D-crystallization including the formation of one metastable
($3~\times~3)-Ni(111$) phase was suggested.

The obtained results shed a light on the growth of the eutectic structure with
nanoscale faceted crystals and can be useful for the creation of nanostructures
based on full or partial crystallization of metallic glasses.

\section{Acknowdledjement}
This work was partially supported by RFBR grants (in particular 13-02-00279-a, 12-02-00206-a), Ministry of Science grants (MK-2780.2013.2) and by World Premier
International  Research Center Initiative (WPI), MEXT, Japan. 



%% References


\bibliographystyle{elsarticle-num}
\bibliography{ArXiv_2013}






% \end{thebibliography}

\end{document}